\documentclass{article}
\usepackage{ijcai26}

\usepackage{times}
\usepackage{soul}
\usepackage{url}
\usepackage[hidelinks]{hyperref}
\usepackage[utf8]{inputenc}
\usepackage[small]{caption}
\usepackage{graphicx} 
\DeclareGraphicsExtensions{.pdf,.png,.jpg} 
\usepackage{amsmath}
\usepackage{amsthm}
\usepackage{booktabs}
\usepackage{algorithm}
\usepackage{algorithmic}
\usepackage[switch]{lineno}
\usepackage{amssymb}

\title{RefiningGPT: Specialized language Models for Automated Refinery Unit-level Process Diagram Synthesis}

\author{
Dongxiao Liu$^{1,2}$\and
Yuwen Ding$^2$\and
Xinghai Wei$^1$\and
Jiacheng Ji$^2$\and
Lei Li$^3$\and
Linghui Li$^1$\and
Xiaoyong Li$^1$
\affiliations
$^1$Beijing University of Posts and Telecommunications, Beijing, China\\
$^2$Sinopec Engineering Incorporation, Beijing, China\\
$^3$Sinopec, Beijing, China
\emails
\{liudongxiao, lilinghui, juntvt, lixiaoyong\}@bupt.edu.cn,
\\
\{liudongxiao.sei, dingyuwen.sei, jijiacheng.sei, lileig.sei\}@sinopec.com
}

\begin{document}

\maketitle

\begin{abstract}
Applying LLMs to complex industrial processes remains challenging due to the semantic gap between natural language design intents and the rigorous physical logic of engineering. 
In the field of petroleum refining engineering, a critical bottleneck is the automated synthesis of Unit-level Process Diagrams (UPDs), which serve as the topological bridge connecting abstract requirements to concrete unit operations. 
In this paper, we propose RefineGPT, a domain-specialized agent for autonomous refinery design.
RefineGPT adopts a hierarchical architecture in which a supervised fine-tuned small language model is responsible for selecting units that satisfy design requirements, while a large language model is used to connect these units to generate the final topology. To enable supervised training, we develop a pipeline that extracts latent process motifs from noisy, unstructured legacy topologies and synthesizes high-quality rationale-based Chain-of-Thought (CoT) training data.
Empirical validation demonstrates that RefineGPT achieves substantial improvements in topological consistency and chemical engineering feasibility, establishing a high-fidelity pathway for AI-augmented industrial process synthesis.


\end{abstract}

\begin{figure}[t]
\centering
\includegraphics[width=1.0\columnwidth]{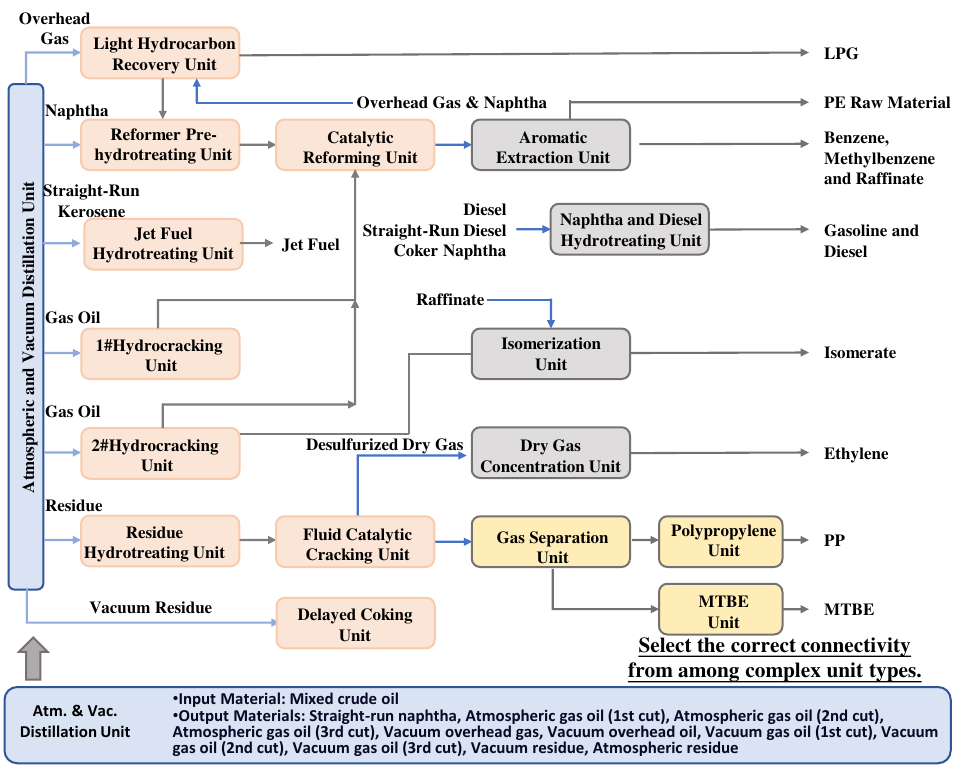} 
\caption{A representative refinery process , illustrating the complex connectivity among unit operations for converting mixed crude oil into diverse petrochemical products (e.g., LPG, gasoline, diesel, ethylene, PP, MTBE). This diagram exemplifies the challenge of automated UPD synthesis: aligning high-level process intent with rigorous engineering motifs under multi-scale material flows.}
\end{figure}

\section{Introduction}

Artificial Intelligence (AI) is reshaping the field of industrial engineering with unprecedented depth and breadth, particularly in the design, optimization, and automated decision-making of complex process flows \cite{cao2022ai,mattioli2023ai}. Large Language Models (LLMs), leveraging their powerful capabilities in natural language understanding and generation, have achieved significant breakthroughs in cross-modal reasoning, knowledge extraction, and instruction following. These advances offer a novel paradigm for transforming human expert experience into computable and executable industrial knowledge \cite{spijkman2025llm,chen2024systems,kampelopoulos2025review}.

However, the application of LLMs in industrial process modeling and flow synthesis remains in its early exploratory stage. Although existing research has attempted to employ language models for molecular design \cite{wang2025llm}, biological reaction path prediction \cite{fallahpour2025bioreason,ko2025reactionreasoner}, or small-scale chemical synthesis route planning \cite{zhang2025synask,zhang2025chemactor,Zhang2024ChemLLMAC}, these efforts largely focus on microscopic scales or specific disciplinary contexts. They have not systematically addressed the critical challenge in large-scale continuous industrial production scenarios (e.g., refining, petrochemicals): how to achieve precise semantic alignment between structured engineering knowledge — such as ``unit operations–material flows–connection topology'' — and natural language descriptions at the macro-process level. This gap severely constrains the transition of LLMs from ``text understanders'' to ``process designers.''

To address this, we propose \textbf{RefiningGPT} — the first specialized framework dedicated to the autonomous synthesis of unit-level refinery topologies. We recognize that directly prompting LLMs to generate structured graphs often leads to semantic drift and logical inconsistency. Therefore, we introduce an innovative ``Think-then-Draw'' intelligent design paradigm: guiding the model to first perform structured engineering reasoning in natural language (Thinking), and then translate this rigorous internal reasoning into a precise graph structure output (Drawing).

Our key contributions are embodied in the following two technical aspects:
\begin{itemize}

    \item \textbf{Domain-adaptive fine-tuning based on process graph topology:} 
    We construct a method to automatically extract ``unit operation--material flow--connection relationship'' topological structures from 20 real-world industrial process diagrams, converting them into a high-quality Supervised Fine-Tuning (SFT) dataset of 500 high-fidelity triplets. Based on this, we perform domain-adaptive fine-tuning on an 8B-parameter small language model (SLM), enabling it to internalize the intrinsic logic and expression norms of refinery engineering, thereby grounding its capabilities in rigorous industrial principles.

    \item \textbf{Hierarchical ``Think-then-Draw'' mechanism with Constraint-Aware RAG:} 
    We design a two-stage architecture where the fine-tuned SLM first selects units and generates an engineering rationale. Subsequently, a knowledge-augmented LLM (e.g., DeepSeek-V3) leverages a specialized Constraint-Aware Retrieval-Augmented Generation (RAG) mechanism to synthesize the final topolog. This process is governed by a hard engineering consistency constraint $\mathcal{C}(G)$ to ensure that every material flow connection complies with physical and functional compatibility.
\end{itemize}

To the best of our knowledge, RefiningGPT is the first specialized framework dedicated to the autonomous synthesis of refinery-level topologies. By grounding generation in a refinery-specific knowledge base, our approach addresses the unique challenge of aligning macro-process descriptions with structured engineering motifs. To validate the effectiveness and generalization of this framework, we comprehensively evaluate RefiningGPT using a two-stage evaluation protocol on ChemFlow-Bench, a multi-scale benchmark derived from twenty representative refineries. To assess generalization across diverse industrial scenarios, the evaluation is conducted on three distinct real-world refinery archetypes: \textit{fuel-type}, \textit{petrochemical-type}, and \textit{aromatics-oriented}. Experimental results demonstrate that RefiningGPT achieves substantial improvements in topological consistency and chemical engineering feasibility, validating its superiority over existing methods and confirming its practical value for real-world refinery design.

\begin{figure*}[!ht]
\centering
\includegraphics[width=0.8\textwidth]{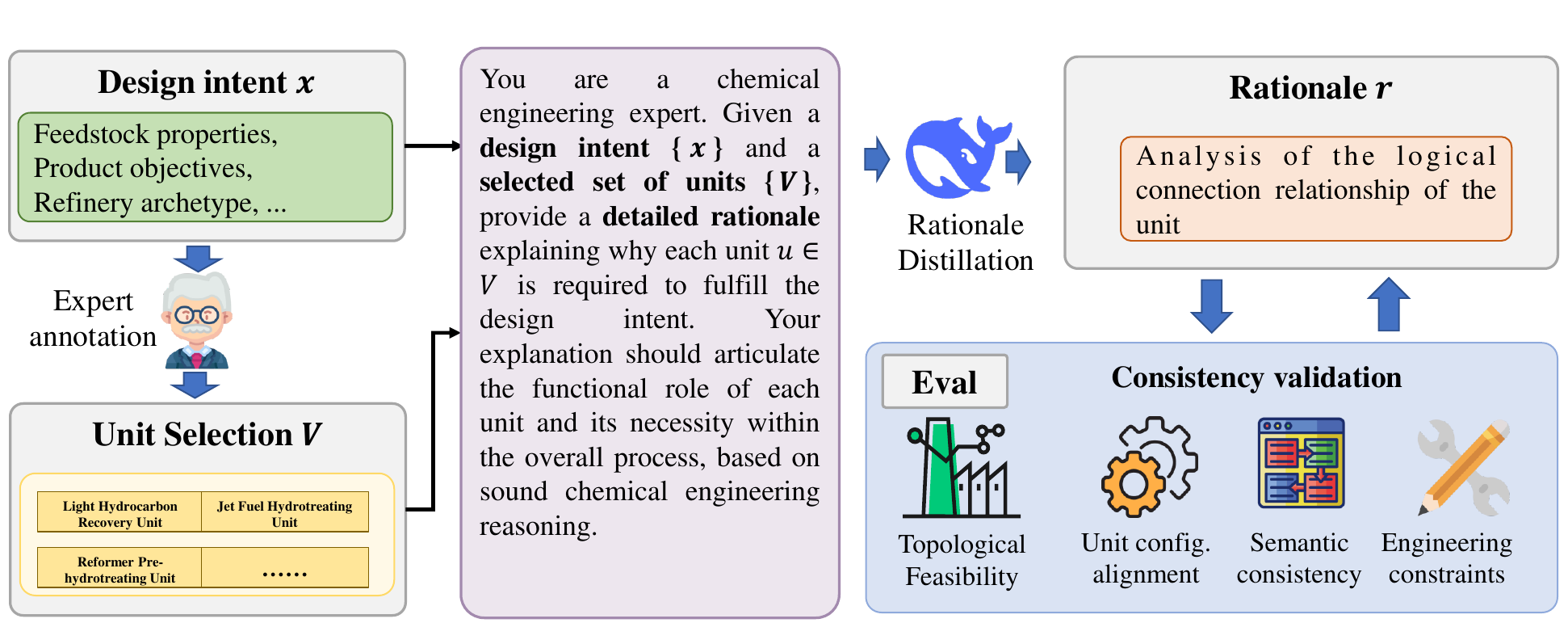} 
\caption{\textbf{Supervised Fine-Tuning (SFT) Dataset Generation Pipeline.} This diagram illustrates the process for automatically extracting and validating high-quality training data from legacy refinery diagrams. The pipeline involves two main stages: (1) \textbf{Rationale Distillation}, where a teacher model generates an engineering rationale ($r$) explaining why specific units ($V$) are selected for a given design intent ($x$); and (2) \textbf{Consistency-aware Validation}, where the generated triplets $(x, r, V)$ undergo a rigorous three-part verification loop to ensure empirical validity, semantic-unit consistency, and adherence to chemical engineering principles before being added to the final SFT dataset.}
\label{fig:rationale_distillation}
\end{figure*}

\section{Related Work}
\paragraph{Chemical Foundation Models and Tool Integration.} The application of large language models (LLMs) to scientific domains has been accelerated by the development of domain-specific benchmarks and foundation models. Early efforts like SciBench~\cite{lu2022learn} and ChemData~\cite{Zhang2024ChemLLMAC} established evaluation protocols for scientific reasoning, paving the way for chemistry-focused models. ChemDFM~\cite{zhao2024chemdfm}, trained on 34B chemical-domain tokens, represents a step toward chemical general intelligence. To further refine the model's grasp of chemical transformations, ReactGPT~\cite{chen2025reactgpt} employs in-context tuning to improve the understanding and prediction of complex chemical reactions. Building on these, ChemCrow~\cite{m2024augmenting} integrates LLMs with expert tools—such as ASKCOS~\cite{tu2025askcos} for retrosynthesis and MolBERT~\cite{fabian2020molecularrepresentationlearninglanguage} for property prediction—to automate multi-step chemical workflows.

\paragraph{LLMs for Automated Industrial Process Design.} While the aforementioned works excel at the molecular or reaction scale, industrial process design requires reasoning over continuous, plant-scale systems governed by mass/energy balances and control logic. Moving toward these industrial applications, Tian et al.~\cite{tian2026textsimulationmultiagentllm} recently proposed a multi-agent workflow to automate chemical process design by bridging natural language descriptions with technical simulations. Although such efforts and synthetic datasets like SynDIP~\cite{srinivas2025autochemschematicaiagenticphysicsaware} aim to teach LLMs unit-operation semantics, the intricate topological complexity and rigorous physical constraints of real-world refineries remain partially addressed in current generative frameworks. Our work bridges this gap by grounding LLM generation in industrial-scale topological constraints.

\paragraph{Reasoning Paradigms and Structured Generation.} Generating structured outputs, such as graphs or process flowsheets, from LLMs often suffers from logical inconsistencies and semantic drift. To mitigate this, the chain-of-thought (CoT) prompting paradigm~\cite{wei2022chain} has proven effective in improving multi-step reasoning by decomposing complex tasks into intermediate steps. Extensions like self-consistency~\cite{wang2023selfconsistencyimproveschainthought} and reflexion~\cite{shinn2023reflexion} further enhance reliability through iterative self-evaluation. Inspired by these advances, we introduce a "Think-then-Draw" paradigm. Unlike general reasoning tasks, our work leverages rationale distillation to enable a specialized small language model to internalize expert engineering logic, generating a rigorous "design rationale" that grounds the subsequent topology synthesis.

\section{Problem Formalization: Intent-Driven Process Synthesis}
We formulate unit-level process diagram (UPD) synthesis as an engineering-constrained directed graph generation problem. Given a design intent $x=(c,p,t,s)$, where $c$ denotes feedstock properties, $p$ denotes product objectives, $t$ denotes the refinery archetype, and $s$ denotes operational and engineering constraints, the objective of UPD synthesis is to generate a directed graph $G=(V,E)$. 
Each node $u \in V$ presents a process unit selected from a unit library $\mathcal{U}$.
The edge set $E \subseteq V \times V$ represents material flow connections between units. 
Each unit $u \in \mathcal{U}$ is characterized by an inherent input material set 
$\text{In}(u) \subseteq \mathcal{M}$ and output material set $\text{Out}(u) \subseteq \mathcal{M}$, where $\mathcal{M}$ denotes the global material property space. To ensure the engineerability of the synthesized topology, a hard engineering consistency constraint $\Phi$ is imposed, defined as:
\begin{equation}
    \Phi(G)=1 \iff \forall (u,v) \in E, \text{Out}(u) \cap \text{In}(v) \neq \emptyset
\end{equation}

This constraint enforces material and functional compatibility along every connection, thereby guaranteeing the structural feasibility of the generated process topology.

\begin{figure*}[!ht]
\centering
\includegraphics[width=0.85\textwidth]{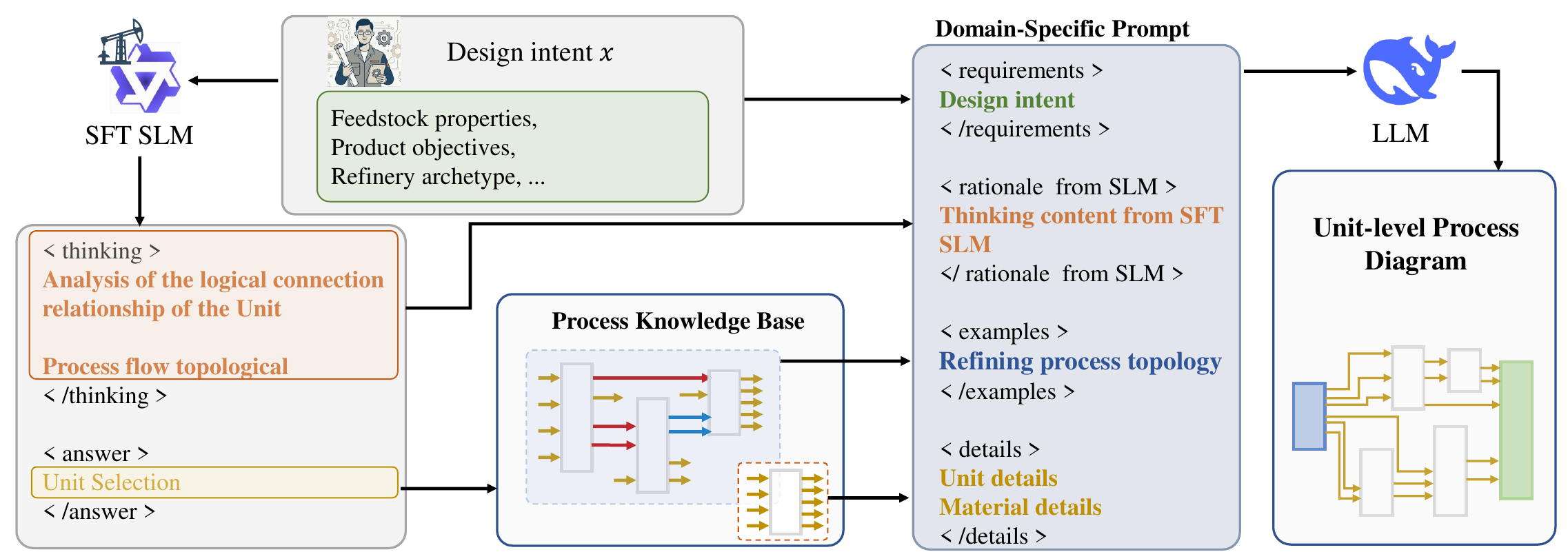} 
\caption{
        \textbf{Overview of the RefiningGPT Framework.}
        The system operates in a two-stage, hierarchical manner.
        First, a domain-specialized small language model (8B parameters) is fine-tuned on rationale-augmented data to perform \emph{unit selection} based on user-specified design intents (e.g., feedstock, product targets).
        It outputs a set of selected units $V$ along with an internal engineering rationale $r$.
        Second, a large, knowledge-augmented LLM (e.g., DeepSeek-V3) receives $V$ and leverages Retrieval-Augmented Generation (RAG) to access a Process Knowledge Base $K$, which contains unit I/O specifications and historical process motifs.
        This enables the LLM to synthesize a functionally complete and physically consistent directed graph $G = (V, E)$ by generating the topology $E$ under strict material compatibility constraints.
    }
\label{fig:overview}
\end{figure*}

\section{Supervised Fine-Tuning Data Generation}
To enable \emph{RefineGPT} to internalize complex refinery engineering logic, we construct a high-quality Supervised Fine-Tuning (SFT) dataset, as illustrated in Figure \ref{fig:rationale_distillation}.
Our goal is to obtain $(x,r,V)$ given the tuple $(x,V)$, where $x$ denotes the design intent, $V$ denotes the selected set of units, and $r$ represents the distilled rationale, which will be used for subsequent fine-tuning of a small language model. The data generation process consists of the following two steps:

\textbf{Rationale distillation.}
Inspired by recent advances in multi-agent reasoning datasets such as ReasonMed~\cite{sun2025reasonmed370kmultiagentgenerated}, we adopt a teacher-student rationale distillation framework to generate interpretable intermediate reasoning processes for training samples. Rationale distillation \cite{wang2024rdrec,wang2025qcrd} allows us to leverage a teacher model (typically a more capable large model) to generate interpretable intermediate reasoning processes for training samples, and then use these reasoning texts as additional supervision signals to train a student model (usually a smaller model), thereby enabling the transfer of reasoning ability.
In RefineGPT, we initialize a set of data pairs $(x,V)$, where $V$ is an expert-annotated set of units that satisfies the given design intent $x$. We then provide these data to the teacher model and prompt it to produce a rationale $r$ explaining why any unit $u \in V$ is necessary for the given design intent $x$.

\textbf{Consistency-aware validation}
To ensure the industrial fidelity of the synthesized data, all triplets $(x, r, V)$ undergo a triple-validation loop as shown in the \texttt{eval} module in Figure 2. Crucially, this validation is not merely a post-hoc check but is deeply integrated into the generation process itself. The teacher model in the Rationale Distillation stage is constrained by the local knowledge base to generate only topologically feasible unit configurations and rationales. Samples failing any check are iteratively refined until physical, logical, and topological closure is achieved.
\begin{itemize}
    \item \textbf{Topological Feasibility:} During Rationale Distillation, the teacher model is constrained by a local knowledge base to generate only topologically valid $V$ and $r$, preventing structurally impossible designs.
    \item \textbf{Unit Configuration Alignment:} The selected unit set $V$ is strictly verified against a local process knowledge base derived from twenty full-scale real-world refinery configurations to ensure empirical validity.
    \item \textbf{Semantic-Unit Consistency:} We verify that every unit mentioned in the rationale $r$ exists in the final unit list $V$, ensuring that the reasoning directly supports the selection.
    \item \textbf{Engineering Mechanism Review:} A secondary logic layer evaluates the CoT against predefined chemical engineering predicates (e.g., ensuring hydrogen balance for hydroprocessing units). 
\end{itemize}

Additionally, we intentionally incorporate negative samples, comprising approximately $10\%$ of the dataset[cite: 18] to enhance the model's critical reasoning and error-detection capabilities. These samples feature perturbed configurations, such as missing critical supporting units. The model is supposed to identify these flaws within its rationale phase and propose corrective measures. The final SFT dataset comprises 500 high-fidelity triplets, establishing a robust foundation for the primary unit selection task in \emph{RefineGPT}.

\section{RefiningGPT Design}

\subsection{Overview}
Figure \ref{fig:overview} illustrates the overview of RefiningGPT, which consists of two key components.
\textit{(i)} A fine-tuned small language model equipped with domain knowledge in petroleum refining. This model takes user-specified design requirements as input and produces coarse-grained selections of units, ensuring that the user requirements are adequately satisfied.
\textit{(ii)} A knowledge-augmented LLM that receives the unit selections from the small language model and leverages external retrieval-augmented generation (RAG) to obtain detailed unit information, thereby generating the final topology.

\subsection{Domain-Aligned Supervised Fine-Tuning}
Given a specific design intent, the fine-tuned model is employed to select the required units, and its reasoning process is further leveraged for downstream topology generation.
To align the model's capabilities with the rigorous logic of refinery engineering, we employ a standard Supervised Fine-Tuning (SFT) strategy on a 8B-parameter transformer backbone\cite{NEURIPS2022_b1efde53}. 
We leverage the high-quality, rationale-augmented dataset described in Section 4 to enforce the ``Think-then-Draw'' paradigm through expert-annotated data design.
For each training triplet $(x, r, V) \in \mathcal{D}$, the model is trained to minimize the standard auto-regressive language modeling objective:

\begin{equation}
\mathcal{L}(\theta) = -\sum_{t=1}^{T} \log P(y_t | y_{<t}, x; \theta),
\end{equation}
where $x$ represents the design intent, and $y = \{r, V\}$ is the concatenated sequence of the engineering rationale $r$ and the unit list $V$. By interleaving the step-by-step reasoning $r$ within \texttt{<thinking>} tags before the final unit selections $V$, the model learns to treat the rationale as a latent state that conditions the final decision. This approach effectively eliminates the need for custom loss functions or manual hyperparameter balancing, while ensuring that unit selections are grounded in physical logic and material balance requirements.

\subsection{Topology Synthesis via Constraint-Aware RAG}
\label{subsec:topology_synthesis}
The units selected by the SLM are provided to the LLM to generate the final topology. To ensure compliance with engineering constraints in the final topology,
we employ a Constraint-Aware Retrieval-Augmented Generation (RAG) mechanism, which grounds the generation process in empirical chemical engineering knowledge:

\paragraph{Schema and Prototype Retrieval} 
For the predicted unit set  $ V $ , the system retrieves precise Input/Output (I/O) specifications for each unit from a verified Process Knowledge Base $ \mathcal{K} $ . This database includes not only rigid I/O definitions but also historical ``thinking chains'' and proven process motifs extracted from real-world refinery configurations.

\paragraph{Contextual Grounding \& Graph Synthesis}
These retrieved engineering specifications are injected into the prompt as \textit{hard constraints}. A powerful LLM (e.g., DeepSeek-V3) is then instructed to generate the edge set  $ E $  by connecting units in  $ V $ . The generation is explicitly conditioned on ensuring that for every proposed edge  $ (u, v) $ , the output material set of unit  $ u $  ( $ \mathrm{Out}(u) $ ) has a non-empty intersection with the input material set of unit  $ v $  ( $ \mathrm{In}(v) $ ), i.e.,  $ \mathrm{Out}(u) \cap \mathrm{In}(v) \neq \emptyset $ . This enforces material and functional compatibility at every connection point.
ially sound design.
To guarantee strict adherence, we combine in-context exemplars demonstrating valid connections with a post-hoc validation loop that iteratively corrects any constraint-violating edges until all connections satisfy the I/O compatibility condition.
This RAG-enhanced approach effectively bridges the gap between the high-level unit selection and the low-level topological construction, ensuring the generated process flowsheet adheres to fundamental engineering principles.

\section{Experiments}
In this section, we present the experimental setup, the benchmark, main results and ablation study.

\subsection{Setup}
To ensure data privacy and deployment security, all models are privately hosted within our internal infrastructure. We experiment with multiple large language models, including Qwen3-8B\cite{yang2025qwen3technicalreport}, Qwen3-32B, Llama3.1-8B\cite{llama3_1_2024}, DeepSeek-V3, DeepSeek-R1 and Qwen3-235B-A22B-Thinking-2507 (accessed via a private API).
For efficient fine-tuning, we adopt the Unsloth framework, which enables high-speed LoRA training with reduced memory overhead. All experiments are conducted on 2× NVIDIA A800-80G GPUs. We use the AdamW optimizer with a learning rate of 5e-5, a per-device batch size of 4, and a maximum input sequence length of 4096 tokens. During inference, we employ a decoding strategy with temperature = 0.3, top-p = 0.95, and top-k = 5.

\subsection{Benchmark}
\label{sec:benchmark}
To evaluate RefiningGTP, we construct ChemFlow-Bench, a multi-scale benchmark derived from real-world refinery configurations.
ChemFlow-Bench is constructed from operational data of twenty representative refineries. 
Unlike general graph datasets, ChemFlow-Bench focuses on the directed flow of materials between specialized chemical processing units. For each task, we provide two levels of ground-truth annotation:

\begin{itemize}
    \item \textbf{Level 1: Unit Selection Ground Truth} \\
    This ground truth annotation specifies the selection of processing units tailored to the given design requirements, including:
    \begin{itemize}
        \item \textit{Unit names}: Each $u \in \mathcal{U}$ is a canonical processing unit name (e.g., crude distillation unit, delayed coking unit),
        \item \textit{Chain-of-Thought rationale}: An expert-written justification grounded in chemical engineering principles (e.g., coking is adopted to manage heavy residues exhibiting high carbon-forming tendency).
    \end{itemize}
    This level enables direct evaluation of unit name accuracy and reasoning quality.

    \item \textbf{Level 2: Unit-level Process Diagram Ground Truth} \\
    This ground truth annotation provides domain-specific topological and material rules, including:
    \begin{itemize}
        \item \textit{Local I/O validity}: Each unit must satisfy its required input/output configuration (e.g., a hydrogenation unit requires hydrogen feed),
        \item \textit{Global material continuity constraints}: Certain streams must be topologically connected to designated treatment or recovery units (e.g., all sour streams must reach a sulfur recovery unit).
    \end{itemize}
 This level supports end-to-end assessment of diagram synthesis fidelity.
\end{itemize}

\subsubsection{Dataset Statistics by Refinery Type}
To support fine-grained evaluation of generalization, we select three real-world refinery configurations from \emph{ChemFlow-Bench} as representative test instances, each corresponding to a distinct operational archetype:
\begin{itemize}
    \item \textbf{Fuel-type}: Optimized for transportation fuels (e.g., gasoline, diesel); features a relatively linear flow with minimal recycling.
    \item \textbf{Petrochemical-type}: Maximizes light olefins (ethylene, propylene) via deep conversion units (e.g., FCC, steam crackers).
    \item \textbf{Aromatics-oriented}: Emphasizes benzene, toluene, and xylene (BTX) production through catalytic reforming and extraction.
\end{itemize}
These instances reflect fundamental differences in process complexity and operational focus, and serve as the basis for our per-type performance analysis.

As summarized in Table~\ref{tab:bench_stats}, the selected refineries contain 11, 22, and 24 distinct processing units, respectively, with 81, 152, and 148 material flows—each modeled as a directed edge in the process graph.

\begin{table}[htbp]
\centering
\small
\begin{tabular}{lcc}
\toprule
\textbf{Refinery Type} & \textbf{Units} & \textbf{Material Flows} \\
\midrule
Fuel-type          & 11 & 81 \\
Petrochemical-type & 22 & 152 \\
Aromatics-oriented & 24 & 148 \\
\bottomrule
\end{tabular}
\caption{Statistics of the three real-world refinery instances used for evaluation in \emph{ChemFlow-Bench}. ``Material Flows'' denote directed edges in the UDP.}
\label{tab:bench_stats}
\end{table}

\paragraph{Data Integrity and Partitioning} To ensure a rigorous evaluation, we implement a strict zero-overlap policy between the SFT training dataset and \emph{ChemFlow-Bench}. While both sets are derived from common refinery engineering principles, the specific process configurations and topological sequences in \emph{ChemFlow-Bench} are systematically excluded from the training corpus. This mutual exclusivity prevents data leakage and ensures that the model's performance on the benchmark accurately reflects its ability to generalize to unseen industrial scenarios.

\begin{table*}[t]
\centering
\scriptsize 
\renewcommand{\arraystretch}{0.78} 
\resizebox{1\linewidth}{!}{
\begin{tabular}{lcccccccc}
\toprule
& \multicolumn{2}{c}{\textbf{Overall}} & \multicolumn{2}{c}{\textbf{Fuel}} & \multicolumn{2}{c}{\textbf{Petrochem}} & \multicolumn{2}{c}{\textbf{Aromatics}} \\
\cmidrule(lr){2-3} \cmidrule(lr){4-5} \cmidrule(lr){6-7} \cmidrule(lr){8-9}
\textbf{Model} & UNF$_1$ & CoT-C & UNF$_1$ & CoT-C & UNF$_1$ & CoT-C & UNF$_1$ & CoT-C \\
\midrule
Llama3.1-8B & 0.40 & 0.21 & 0.42 & 0.25 & 0.34 & 0.13 & 0.43 & 0.24 \\
Qwen3-8B    & 0.43 & 0.32 & 0.57 & 0.32 & 0.38 & 0.36 & 0.36 & 0.29 \\
Qwen3-32B   & 0.40 & 0.35 & 0.46 & 0.31 & 0.27 & 0.43 & 0.47 & 0.32 \\
Qwen3-235B  & 0.30 & 0.40 & 0.32 & 0.39 & 0.41 & 0.48 & 0.48 & 0.34 \\
DeepSeek-V3 & 0.51 & 0.38 & 0.47 & 0.36 & 0.54 & 0.44 & 0.54 & 0.33 \\
DeepSeek-R1 & 0.55 & 0.41 & 0.54 & 0.45 & 0.63 & 0.43 & 0.48 & 0.36 \\
\midrule
\textbf{RefiningGPT (Ours)} & \textbf{0.70} & \textbf{0.68} & \textbf{0.70} & \textbf{0.76} & \textbf{0.72} & \textbf{0.69} & \textbf{0.69} & \textbf{0.63} \\
\bottomrule
\end{tabular}
}

\caption{Stage 1 results: Unit selection ($\text{UNF}_1$) and reasoning quality (CoT-C) across refinery types. All scores are normalized to the range [0,1], with higher values indicating better performance. Best results are marked in \textbf{bold}.
}
\label{tab:stage1}
\end{table*}

\begin{table*}[t]
\centering
\setlength{\tabcolsep}{2.5pt}

\scalebox{1}{
\begin{tabular}{l ccc ccc ccc ccc}
\toprule
\textbf{Model}
& \multicolumn{3}{c}{\textbf{Overall}} 
& \multicolumn{3}{{c}}{\textbf{Fuel}} 
& \multicolumn{3}{c}{\textbf{Petrochem}} 
& \multicolumn{3}{c}{\textbf{Aromatics}} \\
\cmidrule(lr){2-4} \cmidrule(lr){5-7} \cmidrule(lr){8-10} \cmidrule(lr){11-13}
& nGED $\downarrow$ & CSPC $\uparrow$ & IOV $\uparrow$ & nGED $\downarrow$ & CSPC $\uparrow$ & IOV $\uparrow$ & nGED $\downarrow$ & CSPC $\uparrow$ & IOV $\uparrow$ & nGED $\downarrow$ & CSPC $\uparrow$ & IOV $\uparrow$ \\
\midrule
Llama3.1-8B     & 0.61 & 0.40 & 0.34 & 0.62 & 0.41 & 0.34 & 0.64 & 0.38 & 0.32 & 0.58 & 0.40 & 0.35 \\
Qwen3-8B        & 0.67 & 0.32 & 0.33 & 0.68 & 0.32 & 0.33 & 0.70 & 0.30 & 0.31 & 0.64 & 0.33 & 0.34 \\
Qwen3-32B       & 0.70 & 0.33 & 0.31 & 0.71 & 0.33 & 0.31 & 0.73 & 0.31 & 0.29 & 0.67 & 0.34 & 0.32 \\
Qwen3-235B      & 0.65 & 0.35 & 0.28 & 0.66 & 0.35 & 0.28 & 0.68 & 0.33 & 0.26 & 0.62 & 0.36 & 0.29 \\
DeepSeek-V3     & 0.51 & 0.51 & 0.38 & 0.51 & 0.52 & 0.38 & 0.53 & 0.49 & 0.36 & 0.48 & 0.53 & 0.39 \\
DeepSeek-R1     & 0.52 & 0.48 & \textbf{0.44} & 0.52 & 0.48 & 0.46 & 0.54 & 0.46 & 0.44 & 0.49 & 0.49 & \textbf{0.43} \\
\midrule
\textbf{RefiningGPT (Ours)} 
                & \textbf{0.40} & \textbf{0.60} & 0.43 
                & \textbf{0.37} & \textbf{0.61} & \textbf{0.43} 
                & \textbf{0.34} & \textbf{0.63} & \textbf{0.45}
                & \textbf{0.39} & \textbf{0.57} & 0.42 \\
\bottomrule
\end{tabular}
}
\caption{Stage 2 results: Diagram synthesis fidelity across four domains. Metrics: nGED (normalized Graph Edit Distance, lower better), CSPC and IOV (higher better). All scores are normalized to [0,1]. Best results are marked in \textbf{bold}.
}
\label{tab:stage2}
\end{table*}

\subsection{Evaluation Protocol and Metrics}

We evaluate performance in two stages, aligning with our pipeline.

\subsubsection{Stage 1: Unit Selection and Reasoning Quality}
Given an input specification, a model generates a unit sequence $V_{\text{pred}}$ and a CoT justification. The ground-truth unit selection with this input is denoted as $V_{\text{gt}}$. We assess:

\begin{itemize}
    \item \textbf{Unit Selection F1 ($\text{UNF}_1$)}: 
    This metric measures the degree of alignment between the model’s unit selection and the ground-truth solution, and is defined as:
    \begin{equation}
    \notag
        UNF_1=2 \times |V_{\text{pred}} \cap V_{\text{gt}}| / (|V_{\text{pred}}| + |V_{\text{gt}}|)
    \end{equation}
    
\item \textbf{Chain-of-Thought Correctness (CoT-C)}: 
For each correctly selected unit $u \in V_{\text{pred}} \cap V_{\text{gt}}$, domain experts judge whether the CoT provides a chemically sound and non-hallucinated rationale. 
The CoT-C score is the fraction of units in $V_{\text{pred}}$ that are correctly justified, namely:
\begin{equation}
\notag
    \text{CoT-C} = \frac{1}{|V_{\text{pred}} \cap V_{\text{gt}}|} \sum_{u \in |V_{\text{pred}} \cap V_{\text{gt}}|} \mathbb{I}[\text{justification for } u \text{ is valid}].
\end{equation}


\end{itemize}

\subsubsection{Stage 2: Process Diagram Synthesis}
Given the selected unit set $V_{\text{pred}}$, RefineGPT generates a full process diagram $G_{\text{gen}}$. We evaluate  $ G_{\text{gen}} $  against the ground-truth graph  $ G_{\text{gt}} $  using three complementary metrics:

\begin{itemize}

    \item \textbf{Normalized Graph Edit Distance (nGED)}: Measures the structural similarity between the generated graph $ G_{\text{gen}} $ and the ground-truth graph $ G_{\text{gt}} $, accounting for node/edge insertions, deletions, and substitutions. Due to the NP-hard nature of exact GED computation, we employ an approximate algorithm with a predefined upper bound on the search space to ensure computational feasibility while maintaining a reasonable approximation of topological accuracy. The final score is normalized by the maximum number of nodes in either graph, yielding a value in $[0, 1]$, where lower values indicate higher similarity.
    
    \item \textbf{Constraint-Satisfying Path Coverage (CSPC)}: Evaluates whether all critical material routing paths are correctly realized in  $ G_{\text{gen}} $ . For each required path, we check if there exists a valid directed path in  $ G_{\text{gen}} $ . The CSPC score is computed as:

$$
    \text{CSPC} = \frac{\text{Number of satisfied critical paths}}{\text{Total number of required critical paths}}
$$

    \item \textbf{Unit I/O Validity Rate (IOV)}: Assesses the correctness of input/output connections for each unit in  $G_{\text{gen}}$. For each unit  $ u \in \mathcal{U}_{\text{pred}} $, we verify if its incoming/outgoing edges satisfy domain-specific requirements. The IOV score is given by:

$$
    \text{IOV} = \frac{\text{Number of units with valid I/O connections}}{\text{Total number of units in } \mathcal{U}_{\text{pred}}}
$$
 
\end{itemize}

These metrics collectively assess whether improved unit selection and reasoning (Stage 1) translate into more chemically plausible and topologically accurate process diagrams.

\subsection{Main Results}

We evaluate all models on \emph{ChemFlow-Bench} across the three refinery archetypes defined in Section~\ref{sec:benchmark}, reporting performance on both unit selection ($\text{UNF}_1$) and reasoning quality (CoT-C) in Stage 1, and diagram synthesis fidelity in Stage 2.

\paragraph{Stage 1: Unit Selection and Reasoning.}

Table~\ref{tab:stage1} shows that model scale does not guarantee performance: Qwen3-235B (30.0\% $\text{UNF}_1$) underperforms smaller models like Llama3.1-8B (40.0\% $\text{UNF}_1$). In stark contrast, \textbf{RefiningGPT (Ours) achieves state-of-the-art $\text{UNF}_1$ across all domains} (70.0\% overall), outperforming the best baseline (DeepSeek-R1 at 55.0\%) by 15.0 percentage points. This highlights the critical role of domain-specific fine-tuning over raw scale. Crucially, \textbf{RefiningGPT also demonstrates significantly superior reasoning quality}, achieving a CoT-C score of 0.68 overall, substantially higher than the best baseline (DeepSeek-R1 at 0.41). While all models face challenges in perfect reasoning, RefiningGPT's high CoT-C scores indicate its ability to generate more coherent and logically sound justifications for unit selection.

\paragraph{Stage 2: Process Diagram Synthesis.}
Table~\ref{tab:stage2} reveals a key insight: \textit{accurate unit selection does not ensure correct topology}. Despite its strong Stage 1 performance, RefiningGPT must still synthesize valid graphs. It achieves state-of-the-art fidelity in 10/12 metric-domain combinations, notably excelling in the complex Petrochemical domain (CSPC=0.63). Its success stems from generating more units while maintaining high connection accuracy: although adding units increases the denominator for CSPC/IOV, the higher proportion of correctly connected units leads to a net score gain.

In summary, while large-scale models provide strong reasoning capabilities, \textbf{RefiningGPT demonstrates that targeted fine-tuning with domain structure yields superior practical utility}—enabling smaller models to match or exceed black-box giants in structured industrial tasks.

\subsection{Ablation Study}
\label{sec:ablation}
\paragraph{The Impact of Chain-of-Thought Supervision on Unit Selection}
To assess the benefit of engineering rationales in training, we evaluate three configurations: 
the native \texttt{Qwen3-8B} model (42.87\% $\text{UNF}_1$), 
standard supervised fine-tuning on unit selections alone (\textit{SFT}, 60.92\%), 
and our rationale-augmented approach (\textit{CoT-SFT}, 70.30\%). 
The 9.4 percentage point improvement from \textit{SFT} to \textit{CoT-SFT} demonstrates that explicitly training the model to generate and condition on engineering rationales yields more accurate and physically consistent unit selection.

\paragraph{Quality of Small-Model Reasoning}
We conduct an ablation study to quantify the contribution of different rationale components generated by our compact 8B SFT model. As shown in Table~\ref{tab:ablation_avg}, providing only internal reasoning ($r_{\text{reason}}$) boosts the executor’s average UNF$_1$ to 73.85\% — a +6.95\% gain over the “None” baseline and just 5.3\% below human expert rationales — demonstrating that even small, domain-specialized models can encode sufficient mechanistic knowledge for actionable guidance. Crucially, incorporating \textit{key topology} yields the largest improvement (78.9\% UNF$_1$), while combining both reasoning and topology (\textit{All}) achieves the best performance (83.15\% UNF$_1$), validating our core design: structured topological constraints are paramount for synthesis, with reasoning providing essential contextual grounding.

\paragraph{Impact of Context Composition}

\begin{figure}[ht]
    \centering
    \includegraphics[width=0.45\textwidth]{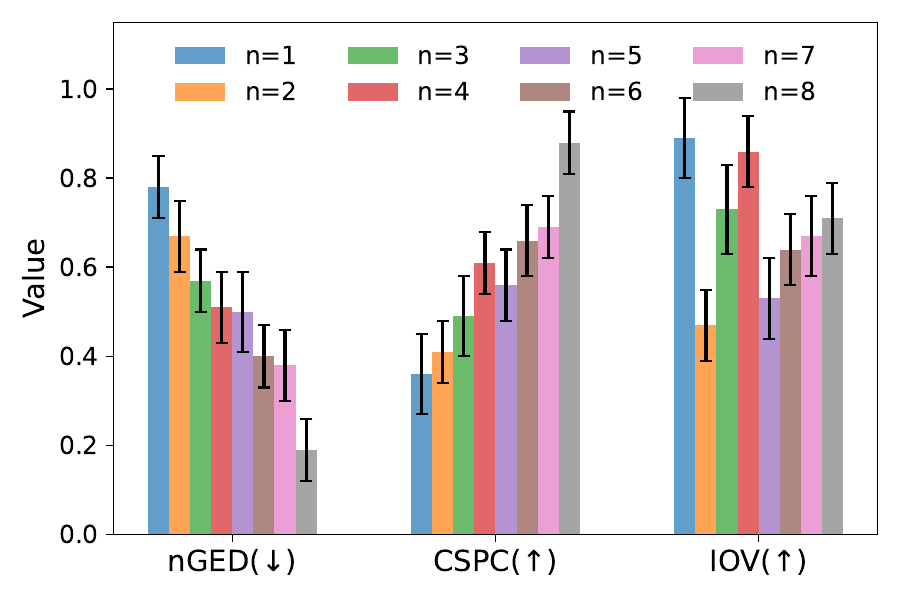}
    \caption{Model performance versus context size $N$. While performance improves up to $N=5$, the sharp drop in nGED at $N=8$ and declining IOV indicate growing instability — not convergence to an optimal balance — as context length increases.}
    \label{fig:context_ablation}
\end{figure}

As shown in Figure~\ref{fig:context_ablation}, the model’s structural fidelity (nGED) and path coverage (CSPC) both improve monotonically as context size $N$ increases from 1 to 5 — confirming that richer historical information enhances topological accuracy and constraint satisfaction. However, this trend does not extend to physical feasibility: the Unit I/O Validity Rate (IOV), which measures whether generated units maintain valid input/output connections, peaks at small $N$ (0.89 at $N=1$) but declines sharply as context expands (e.g., to 0.53 at $N=5$). This stark divergence indicates that while the model benefits structurally from extended context, it becomes increasingly unstable in maintaining domain-specific connectivity constraints — revealing a critical limitation in its ability to generate physically consistent full-scale devices under long-range dependencies. The drop in IOV is not an artifact of poor retrieval, but rather a symptom of the model’s inability to scale constraint enforcement reliably across sequential unit generation.

\begin{table}[t]
\centering
\small
\setlength{\tabcolsep}{3.8 pt}  
\renewcommand{\arraystretch}{1.0}
\begin{tabular}{lccccc}
\toprule
\textbf{Rationale Type} & \textbf{nGED$\downarrow$} & \textbf{CSPC$\uparrow$} & \textbf{IOV$\uparrow$} & \textbf{UNF$_1$$\uparrow$} \\
\midrule
\textbf{None} & 0.40 & 0.67 & 0.71 & 0.67 \\
\textbf{Only Reasoning} & 0.36 & 0.67 & 0.93 & 0.74 \\
\textbf{Only Key Topology} & 0.33 & 0.73 & 0.75 & 0.79 \\
\textbf{Key Topology + Reasoning} & \textbf{0.32} & \textbf{0.73} & \textbf{0.84} & \textbf{0.83} \\
\bottomrule
\end{tabular}
\caption{Ablation study on rationale components for the 8B SFT model. We evaluate the impact of different rationale inputs (none, reasoning only, key topology only, or both) on the executor's performance. }
\label{tab:ablation_avg}
\end{table}

\section{Conclusion}
We present \textbf{RefiningGPT}, the first specialized framework for automated unit-level process diagram synthesis in petroleum refining. By integrating domain-adaptive fine-tuning with a constraint-aware RAG mechanism, it bridges natural language intent and engineering physics through a ``Think-then-Draw'' paradigm. Evaluated on ChemFlow-Bench, RefiningGPT achieves state-of-the-art performance in unit selection, reasoning quality, and topology feasibility — especially under complex petrochemical constraints. Our results demonstrate that targeted knowledge injection, not just model scale, is key to high-fidelity industrial process design.

\newpage
\appendix

\section*{Ethical Statement}

All research methodologies, data usage, and experimental procedures adhere to academic integrity standards, industrial ethical norms of the petroleum refining sector, and data security regulations of collaborating commercial entities. The data used to construct the ChemFlow-Bench benchmark was obtained with explicit authorization and underwent comprehensive anonymization and de-identification to protect commercial privacy and intellectual property. All data processing, model training, and experimental validation were conducted within secure commercial intranet environments, with no risk of sensitive data leakage. In respect of commitments to partner data security, original datasets and sensitive experimental data will not be publicly released or open-sourced.



\bibliographystyle{named}
\bibliography{ijcai26}

\end{document}